\documentclass[12pt]{article}
 \usepackage{epsfig}
 \def\be{\begin{equation}}
 \def\ee{\end{equation}}
 \def\bea{\begin{eqnarray}}
 \def\eea{\end{eqnarray}}
 \usepackage{graphicx}% Include figure files

 \catcode`\@=11
 \def\lsim{\mathrel{\mathpalette\@versim<}}
 \def\gsim{\mathrel{\mathpalette\@versim>}}
 \def\@versim#1#2{\vcenter{\offinterlineskip
 \ialign{$\m@th#1\hfil##\hfil$\crcr#2\crcr\sim\crcr } }}
 \catcode`\@=12

 \catcode`@=12
\begin{document}
 \thispagestyle{empty}
 \begin{flushright}
 UCRHEP-T563\\
 January 2016\
 \end{flushright}
 \vspace{0.6in}
 \begin{center}
 {\LARGE \bf Compendium of Models from\\
 a Gauge U(1) Framework\\}
 \vspace{1.2in}
 {\bf Ernest Ma\\}
 \vspace{0.2in}
 {\sl Physics \& Astronomy Department and Graduate Division,\\ 
 University of California, Riverside, California 92521, USA\\}
 \vspace{0.1in}
 {\sl HKUST Jockey Club Institute for Advanced Study,\\ 
 Hong Kong University of Science and Technology, Hong Kong, China\\}

 \end{center}
 \vspace{1.2in}

\begin{abstract}\
A gauge $U(1)$ framework was established in 2002 to extend the supersymmetric 
standard model.  It has many possible realizations.  Whereas all have the 
necessary and sufficient ingredients to explain the possible 750 GeV 
diphoton excess, observed recently by the ATLAS Collaboration at the 
Large Hadron Collider (LHC), they differ in other essential aspects. 
A compendium of such models is discussed.
\end{abstract}

 \newpage
 \baselineskip 24pt

\section{Introduction}

The recent announcement~\cite{atlas15} by the ATLAS Collaboration 
at the Large Hadron Collider (LHC) of a diphoton excess around 750 GeV 
has excited the high-energy phenomenology community in recent weeks.  
In a short note~\cite{m16}, I have pointed out that a gauge $U(1)$ 
framework I established in 2002~\cite{m02} has exactly all the necessary 
and sufficient particles and interactions for explaining this observation.  
There are actually many explicit realizations of this proposal.  
All contain the ingredients to accommodate the diphoton excess, but they 
differ in other essential aspects, such as neutrino mass, leptoquark, 
or diquark interactions, etc.  This paper discusses each in turn. 
One specific version was already studied in 2010~\cite{m10}.
\begin{table}[htb]
\caption{Particle content of gauge $U(1)$ framework.}
\begin{center}
\begin{tabular}{|c|c|c|c|c|c|}
\hline
Superfield & $SU(3)_C$ & $SU(2)_L$ & $U(1)_Y$ & $U(1)_X: (A)$ & 
$U(1)_X: (B)$ \\
\hline
$Q = (u,d)$ & 3 & 2 & 1/6 & $n_1$ & $n_1$ \\
$u^c$ & $3^*$ & 1 & $-2/3$ & $(7n_1+3n_4)/2$ & $5n_1$ \\
$d^c$ & $3^*$ & 1 & 1/3 & $(7n_1+3n_4)/2$ & $2n_1+3n_4$ \\
\hline
$L = (\nu,e)$ & 1 & 2 & $-1/2$ & $n_4$ & $n_4$ \\
$e^c$ & 1 & 1 & 1 & $(9n_1+n_4)/2$ & $3n_1+2n_4$ \\
$N^c$ & 1 & 1 & 0 & $(9n_1+n_4)/2$ & $6n_1-n_4$ \\  
\hline
$\phi_1$ & 1 & 2 & $-1/2$ & $-3(3n_1+n_4)/2$ & $-3(n_1+n_4)$ \\
$\phi_2$ & 1 & 2 & 1/2 & $-3(3n_1+n_4)/2$ & $-6n_1$ \\
$S_1$ & 1 & 1 & 0 & $-(3n_1+n_4)$ & $-(3n_1+n_4)$ \\
$S_2$ & 1 & 1 & 0 & $-2(3n_1+n_4)$ & $-2(3n_1+n_4)$ \\
$S_3$ & 1 & 1 & 0 & $3(3n_1+n_4)$ & $3(3n_1+n_4)$ \\
\hline
$U$ & 3 & 1 & 2/3 & $-4n_1-2n_4$ & $-6n_1$\\
$D$ & 3 & 1 & $-1/3$ & $-4n_1-2n_4$ & $-6n_1$ \\
$U^c$ & $3^*$ & 1 & $-2/3$ & $-5n_1-n_4$ & $-3(n_1+n_4)$ \\
$D^c$ & $3^*$ & 1 & 1/3 & $-5n_1-n_4$ & $-3(n_1+n_4)$ \\
\hline
\end{tabular}
\end{center}
\end{table}

The particle content of this gauge $U(1)_X$ extension of the supersymmetric 
standard model is fixed.  Whereas certain interactions are mandatory, 
others are not.  As explained in Ref.~\cite{m02}, different models come 
from choosing one of two classes of solutions: (A) or (B).  For each, 
there is also the ratio of two charges which may vary.  Hence there 
are many possible models within this framework.  Each will have all the 
mandatory interactions required to explain the 750 GeV observation, but 
will have different predictions regarding other phenomena.

\section{Generic Solutions of Classes (A) and (B)}

Consider the gauge group $SU(3)_C \times
SU(2)_L \times U(1)_Y \times U(1)_X$ with the particle content of 
Ref.~\cite{m02} as shown in Table 1.  There are three copies of 
$Q,u^c,d^c,L,e^c,N^c,S_1,S_2$; two copies of $U,U^c,S_3$; and one copy of 
$\phi_1,\phi_2,D,D^c$.
The following terms of the superpotential are always allowed:
\begin{eqnarray}
&& Q u^c \phi_2, ~~~ Q d^c \phi_1, ~~~ L e^c \phi_1, ~~~ L N^c \phi_2, ~~~ 
S_3 \phi_1 \phi_2, \\ 
&& S_3 U U^c, ~~~ S_3 D D^c, ~~~ S_1 S_2 S_3.
\end{eqnarray}
The charges $n_1$ and $n_4$ are arbitrary, except that $3n_1+n_4 \neq 0$ 
is required to forbid the $\mu \phi_1 \phi_2$ term of the minimal 
supersymmetric standard model (MSSM).  
\begin{figure}[htb]
\vspace*{-3cm}
\hspace*{-3cm}
\includegraphics[scale=1.0]{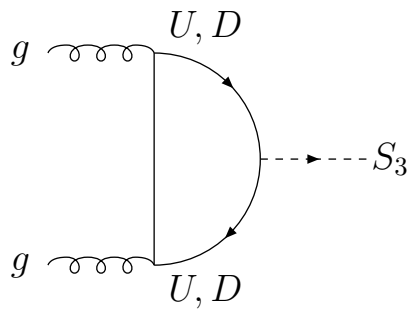}
\vspace*{-21.5cm}
\caption{One-loop production of $S_3$ by gluon fusion.}
\end{figure}
Hence $S_3$ always has the interactions which allow it to be produced by 
gluon fusion in one loop as shown in Fig.~1, and then decays in one 
loop to two photons as shown in Fig.~2.
\begin{figure}[htb]
\vspace*{-3cm}
\hspace*{-3cm}
\includegraphics[scale=1.0]{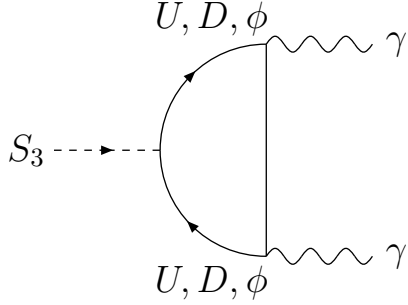}
\vspace*{-21.5cm}
\caption{One-loop decay of $S_3$ to two photons.}
\end{figure}
It may also decay into $S_1 S_2$ final states directly and increase its 
total width.  These are then the essential ingredients which could explain 
the 750 GeV observation.

In choosing $n_1$ and $n_4$, if the resulting model has only those 
interactions of Eqs.~(1) and (2), then the $U,D$ particles are 
stable.  They may form bound states with the known quarks and 
become exotic stable matter.  In the following, only cases 
with additional interactions are considered.

\section{Leptoquark Models}

In (A) for $n_1=0$, the following interactions become allowed:
\begin{equation}
u^c N^c U, ~~~ u^c e^c D, ~~~ d^c N^c D, ~~~ QL D^c, ~~~ N^c N^c S_1.
\end{equation}
This is the case studied in Ref.~\cite{m10} and used in Ref.~\cite{m16} 
for illustration.  Now $U^c,D^c$ should be considered as leptoquark 
superfields, which may also be relevant~\cite{bn15} in understanding 
other possible LHC flavor anomalies.  For $\langle S_1 \rangle \neq 0$, 
$N^c$ acquires a large Majorana mass, hence $\nu$ gets a small Majorana 
seesaw mass in the usual way.

In (B) for $n_4=-n_1$, the following interactions become allowed:
\begin{equation}
u^c e^c D, ~~~ d^c N^c D, ~~~ QL D^c, ~~~ L S_1 \phi_2, 
~~ N^c S_2 S_3.
\end{equation}
Now $D^c$ is a leptoquark, but $U$ is stable because the $U^c D^c D^c$ 
term is not possible as an $SU(3)$ singlet.  Neutrino masses are forced 
to be Dirac.

In (B) for $n_4=5n_1$, the only allowed new interaction is
\begin{equation}
u^c N^c U.
\end{equation}
Hence $U$ is a leptoquark, but $D$ is a stable heavy quark.
Neutrino masses must again be Dirac.

\section{Diquark Models}

In (A) for $n_4=-n_1$, the following interactions become allowed:
\begin{equation}
u^c d^c D^c, ~~~ d^c d^c U^c, ~~~ Q Q D, ~~~ N^c S_2.
\end{equation}
Now both $U^c,D^c$ are diquarks, and neutrinos obtain seesaw Dirac 
masses as follows.  In the space spanned by $(\nu, S_1, N^c, S_2)$, 
the $12 \times 12$ neutrino mass matrix is of the form
\begin{equation}
{\cal M}_\nu = \pmatrix{0 & 0 & m_D & 0 \cr 0 & 0 & 0 & m_S \cr 
m_D & 0 & 0 & M \cr 0 & m_S & M & 0},
\end{equation}
where $m_D$ comes from $\nu N^c \langle \phi_2^0 \rangle$, $m_S$ from 
$S_1 S_2 \langle S_3 \rangle$, and $M$ from $N^c S_2$.  This is thus 
a Dirac seesaw with $m_\nu \simeq m_D m_S/M$.

In (B) for $n_1=0$, the following interactions become allowed:
\begin{equation}
u^c d^c D^c, ~~~ Q Q D.
\end{equation}
Now $D^c$ is a diquark, but $U$ is stable because the $UDD$ term is  
not possible as an $SU(3)$ singlet.  Further, $N^c$ and $S_1$ transform 
in the same way under 
$U(1)_X$, so that a linear combination pairs up with $\nu$ to form 
Dirac neutrinos. 

In (B) for $n_1=-3n_4$, the only allowed new interaction is
\begin{equation}
d^c d^c U^c.
\end{equation}
Hence $U$ is a diquark, but $D$ is a stable heavy quark.
Neutrino masses must again be Dirac.

\section{Heavy Quark Models}

The $U^c,D^c$ singlets may transform in the same way as $u^c,d^c$ under 
$U(1)_X$.  In that case, they will mix and the heavy ones will decay 
to the lighter ones.  Another possibility is that $u^c U$ or $d^c D$ 
is an allowed mass term under $U(1)_X$, in which case there is again 
mixing.

In (A) for $n_4 = -(17/5)n_1$, $U^c,D^c$ and $u^c,d^c$ transform in the 
same way under $U(1)_X$.  In (B) for $n_4 = -(8/3)n_1$, $U^c$ and $u^c$ 
transform in the same way, but $D$ remains stable.  In (B) 
for $n_4 = -(5/6)n_1$, $D^c$ and $d^c$ transform in the same way, 
but $U$ remains stable.  In (B) for $n_4 = (4/3)n_1$, 
$d^c D$ is a mass term, but $U$ is also stable.  In all cases, neutrino 
masses are Dirac.

In (A) for $n_4 = -13n_1$ and in (B) for $n_4 = -(4/3)n_1$, the 
$d^cD^cU^c$ term is allowed.  This means that only one of the exotic 
$U,D$ states is stable.

\section{Majorana Neutrino Mass Models}

To allow Majorana neutrino masses, the term $S_i N^c N^c$ should be present. 
For $S_1 N^c N^c$, it implies $n_1=0$ in (A) and $n_4=3n_1$ in (B). 
For $S_2 N^c N^c$ which automatically allows $S_1 N^c$, it implies 
$n_4=3n_1$ in (A) and $n_4 = (3/2)n_1$ in (B).  For $S_3 N^c N^c$, 
it implies $n_4 = -(9/2)n_1$ in (A) and $n_4 = -21 n_1$ in (B). 
In all cases except the first, i.e. $n_1=0$ in (A) which leads to Eq.~(3), 
the exotic $U,D$ quarks are stable and there is no other interaction 
involving them.

The $N^c N^c$ term by itself is allowed if $n_4 = -9n_1$ in (A) or 
$n_4 = 6n_1$ in (B).  There is however no other allowed term beyond 
Eqs.~(1) and (2).  The exotic $U,D$ quarks are stable in these cases.

\section{Conclusion}

The two most plausible models are those described by Eqs.~(3) and (6). 
The former~\cite{m10} has $U^c,D^c$ as leptoquarks, and neutrino masses 
are Majorana from a TeV scale seesaw mechanism.  The latter has 
$U^c,D^c$ as diquarks, and neutrino masses are Dirac from a high scale 
seesaw mechanism.  In most other models, either $U$ or $D$ or both are 
stable.  Neutrino masses are Dirac in most cases with no 
understanding of why they are so small.

Since the $U(1)_X$ charge assignments of quarks and leptons are all 
different in these various models, the key is in the observation of the 
associated $Z_X$ gauge boson.  If the LHC finds a $Z'$ gauge boson, its 
decay branching fractions~\cite{gm08} would help distinguish among 
possible models of this gauge $U(1)$ framework.

\noindent \underline{\it Acknowledgement}~:~
This work was supported in part by the U.~S.~Department of Energy Grant 
No. DE-SC0008541.

%\newpage
\bibliographystyle{unsrt}

\end{document}